%% file: StronglyInteractingNeutrinoPortalDM-JHEP.tex
\documentclass[letter,11pt]{article}

\pdfoutput=1

\usepackage{jheppub} 
\usepackage{bbm}
\usepackage{bm}
\usepackage{color}
\usepackage{amsmath}
\usepackage{amssymb}
\usepackage{graphicx}
\usepackage{slashed}
\usepackage{eufrak}
\usepackage{bbding}
\usepackage{cleveref}

\input mylatex

\def\hc{\mbox{H.c.}}

\def\phit{\tilde\phi}

\def\pmns{V_{\tt PMNS}}
\def\pmnsd{V^\dagger_{\tt PMNS}}

\def\atm{\acal}
\def\nucl{\ncal}

\def\cw{c_{\tt W}}
\def\sw{s_{\tt W}}
\def\mz{m_{\tt Z}}
\def\mw{m_{\tt W}}
\def\mh{m_{\tt H}}
\def\mv{m_{\tt V}}
\def\mf{m_\fcal}
\def\lx{\lambda_x}
\def\msc{m_\Phi}
\def\mfe{m_\Psi}
\def\mnu{m_\nucl}
\def\ma{m_\atm}
\def\bP{\beta_\Psi}
\def\bv{\beta_{\tt V}}
\def\Lam{m_{\tt N}}
\def\tr#1{\text{tr}\left\{#1\right\}}
\def\r#1#2{r_{{\tt #1}{\tt #2}}}
\def\vh{v_{\tt H}}
\def\prot{{\mathfrak{p}}}
\def\neut{{\mathfrak{n}}}
\def\BS{[\Psi_{\!+}\!\Psi_{\!-}]}

\title{Strongly Interacting Neutrino Portal Dark Matter}

\author[a]{J.M. Lamprea,}
\author[a]{E. Peinado,}
\author[b]{S. Smolenski,}
\author[b]{J. Wudka}

\affiliation[a]{Instituto de F'sica, Universidad Nacional Aut—noma de M\'exico, \\ A.P. 20-364, Ciudad de M\'exico 01000, M\'exico}
\affiliation[b]{Department of Physics and Astronomy, UC Riverside, \\ Riverside, California 92521-0413, USA}

\emailAdd{jmlamprea@estudiantes.fisica.unam.mx}

\emailAdd{epeinado@fisica.unam.mx}

\emailAdd{ssmol001@ucr.edu}

\emailAdd{jose.wudka@ucr.edu}

\abstract{We present a realistic, simple and natural model of strongly-interacting dark matter based on the neutrino-portal paradigm.  The strong interactions at small velocities are generated by the exchange of dark photons, and produce the observed core-like DM distribution in galactic centers; this effect could be spoiled by the formation of DM bound states (also due to dark-photon effects), which we avoid by requiring the DM candidates to be light, with masses below $O(10 \, \gev)$. The mixing of the dark photon with the $Z$ and ordinary photon is strongly suppressed by introducing a softly-broken discrete symmetry similar to charge conjugation, which also ensures that the dark photon life-time is short enough to avoid restrictions derived form big-bang nucleosynthesis and large-scale structure formation. Other constraints are accommodated without the need of fine tuning, in particular nucleon scattering occurs only at one loop, so direct detection cross sections are naturally suppressed. Neutrino masses are generated through the inverse see saw.
}

\begin{document}

\maketitle

\flushbottom

\section{Introduction}

The nature of dark matter (DM) remains one of the most perplexing problems in modern particle and astroparticle physics. Current evidence for the existence of massive particles that interact weakly  with the Standard Model (SM) is entirely gravitational \cite{Zwicky:1933gu,Rubin:1970zza,Corbelli:1999af,Allen:2011zs,Clowe:2006eq}, and every attempt at direct~\cite{Hanany:2019lle,Behnke:2016lsk,Fu:2016ega,Akerib:2016lao,Aprile:2017iyp,Akerib:2016vxi,Tan:2016zwf}, indirect~\cite{Bulbul:2014sua,Ruchayskiy:2015onc,Franse:2016dln,Urban:2014yda,Aharonian:2016gzq,Choi:2015ara,Aartsen:2016zhm,TheFermi-LAT:2017vmf,Hooper:2010mq,Ackermann:2015zua,Cui:2016ppb} or collider~\cite{Aaboud:2016wna,Sirunyan:2018gka,Abercrombie:2015wmb} detection has only led to increasingly stronger constraints on models. In addition, estimations of the DM distribution in dwarf galaxies indicate that the DM density at the core does not exhibit a spike, as would be expected if it behaved as an ideal gas. This 
``core vs. cusp" problem~\cite{deBlok:2009sp,Flores:1994gz,Moore:1994yx,Moore:1999gc} can be alleviated~\cite{Spergel:1999mh,Tulin:2017ara} by including self-interactions within the dark sector; such interactions must be relatively strong and velocity-dependent. Models of this type are often referred to as strongly-interacting dark matter (SIDM) models.

In this paper we will discuss a simple SIDM model that meets all available constraints without fine tuning of parameters. The model is an extension of one discussed earlier~\cite{Gonzalez-Macias:2016vxy,Gonzalez-Macias:2016snw}, based  on the neutrino-portal paradigm~\cite{Cosme:2005sb,Falkowski:2009yz,An:2009vq,Lindner:2010rr,Falkowski:2011xh,Farzan:2011ck,Heeck:2012bz,Baek:2013qwa,Baldes:2015lka,Gonzalez-Macias:2016vxy,Gonzalez-Macias:2016snw,Batell:2017rol,HajiSadeghi:2017zrl,Berlin:2018ztp,Bandyopadhyay:2018qcv,Blennow:2019fhy} where the dark sector couples to the SM via (Dirac) fermion mediators that mix with the SM neutrinos. The dark sector contains two quasi-degenerate fermions, which constitute the relic density, and a scalar, more massive than the fermions. Interactions within the dark sector are mediated by a dark photon, whose mixing with the ordinary photon is (again, naturally) strongly suppressed, since it occurs at three loops; the main decay mode of the dark photon is into neutrinos, and appears at one loop, so the dark photon is relatively long-lived. Of special interest is that the DM self-interactions are useful in suppressing a possible cusp in the DM galactic distribution only when the DM mass is light, below $O( 10)$ GeV.

The paper is organized as follows: in the next section we describe our model~\cite{Gonzalez-Macias:2016vxy} concentrating on the interactions within the dark sector; detailed discussion of the other aspects can be found in the original paper. In sections \ref{sec:ew-constraints}, \ref{sec:rel-ab} and \ref{sec:dd} we discuss the electroweak, relic abundance and direct-detection constraints respectively. Section \ref{sec:numerics} contains results from numerical simulations, and we present our conclusions in section \ref{sec:conclusions}.

\section{The Model}
\label{sec:model}

As noted above, we will study an extension of the neutrino portal dark matter model discussed in~\cite{Gonzalez-Macias:2016vxy}, where we add self-interactions to the dark sector, and double the number of fermions (the justification for this is provided below). The dark sector then contains two fermions $ \Psi_\pm $ with masses $ m_\pm$, and one complex scalar $ \Phi $ with mass $\msc > m_\pm $;  the fermions correspond to the DM. The dark sector is connected to the \sm\ through a set of three (Dirac) neutral fermionic mediators $ \fcal $ with interactions of the form $ \bar\Psi \Phi \fcal $ and~\footnote{$l$ denotes the \sm\ left-handed lepton isodoublet, $\phi$ the Higgs isodoublet and $ \phit = i \sigma_2 \phi^*$, with $ \sigma_2 $ the usual Pauli matrix;  $ l$ and $ \fcal $ carry a family index that we suppress.} $ \bar l \fcal \phit$. 

We generate interactions within the dark sector by assuming the dark sector has a  $ \ui_{\tt dark} $  gauge symmetry under which $ \Psi_\pm$ and $ \Phi $ are charged; we denote by $V$ the corresponding gauge boson, the dark photon. To implement the SIDM paradigm we will assume the $V$ has a non-zero mass $ \mv $ that we introduce using the St\"uckelberg trick. The cross sections generated by $V$ exchange  can then generate  self-interactions  with the  velocity dependence~\cite{Tulin:2017ara}  required to address the core vs. cusp problem. 

Models of this type contain a kinetic mixing term of the form $ \xi V_{\mu\nu} B^{\mu\nu}$ \cite{Holdom:1985ag}, where $B$ is the \sm\ hypercharge gauge field. The coupling $\xi $ is strongly constrained by data: $\xi \le 10^{-3}$ \cite{Alexander:2016aln}; we interpret this as an indication that  the model should contain a symmetry that forbids this interaction and which is either exact or softly broken. For this reason we impose a dark $\mathbb{Z}_{2}$ symmetry (dark charge conjugation -- DCC) under which $V$ is odd and all SM particles are even: the dark scalar has the expected $ \Phi \to \Phi^*$ behavior, while  $ \{ \Psi_+,\, \Psi_-\} $,  form  a dark-charge doublet, exchanged under DCC:
\beq
\mbox{DCC:} \quad \Psi_+ \leftrightarrow  \Psi_-\,, \quad
\Phi\leftrightarrow  \Phi^*\,,\quad
V\leftrightarrow -V \,.
\eeq
The DCC symmetry requires that $\Psi_+$ and $\Psi_-$ have the same mass and couplings; and it also implies that a sufficiently light $V$ will be stable, which is phenomenologically troublesome. For this last reason we will assume that DCC is softly broken by assuming the $ \Psi_\pm $ masses are split (this is the only way to achieve this soft braking with the particle content we assume).   

The Lagrangian for this model is then given by
\bal
\lcal = &\bar\Psi_+ (i\slashed{D}_{+}-m_+)\Psi_+ +\bar\Psi_-(i\slashed{D}_{-}-m_-)\Psi_- +|D\Phi|^{2} \mcr
&-\frac{1}{2}\msc^2|\Phi|^{2}-\frac{1}{4}\lambda|\Phi|^{4}-\frac{1}{4}V_{\mu\nu}V^{\mu\nu}+\half \mv^{2}\left( V_{\mu}-\inv\mv\partial_{\mu}\sigma \right)^{2}+\bar{ \fcal }(i\slashed{\partial}-\mf) \fcal  \mcr
&-\left[\bar{l}Y^{(\nu)} \fcal \tilde{\phi}+ \hc \right]-
\left[ \left( \bar\Psi_+\Phi+\bar\Psi_-\Phi^* \right)(z\fcal)+ \hc \right]-\lambda_{x}|\Phi|^{2}|\phi|^{2}\,,
\label{eq:lag}
\end{align}
where, as noted above, $l$ is the SM left-handed lepton isodoublet and $ \phi$ the SM isodoublet; also
\beq
D^\alpha_\pm=\partial^\alpha \pm i g V^\alpha
\eeq
is the covariant derivative, and
\beq
m_\pm = \mfe \pm \mu \,,
\eeq
where $ \mu $, the fermion mass splitting, parameterizes the soft breaking of DCC; $\sigma$ is the auxiliary field used in the St\"uckelberg trick (the unitary gauge corresponds to $ \sigma =0 $). Finally, we assume three $ \fcal $ fields~\footnote{We have suppressed all family indices.}, hence $ \mf $ and $ Y\up\nu $ are $ 3 \times 3 $ mass and Yukawa coupling matrices, respectively, and $ z $ is a $ 3 \times 1 $ vector. 

Compared to the earlier version, this model has 3 additional parameters: $ \mu,\,g$ and $\mv$. We will see (cf. Sect. \ref{sec:DM-si}), however, that the constraints on the DM self-interactions are sufficient to fix $ \mv$ and $g$ as functions of $ \mfe $, so that, in fact only one additional free parameter is introduced.

Once the \sm\ symmetry is broken the neutrinos $ \nu_L$ (contained in $l$) will mix with the $ \fcal $; we will denote the mass eigenstates as $n_L$, left-handed and massless, and $N$, with a mass of order $ \mf$. To reduce the number of parameters we will assume for simplicity that the $N$ are degenerate, with mass $ \Lam $. In this case the gauge and mass eigenstates are related by
\bal
\fcal &= \ccal N_L + \scal n_L + N_R \,;\mcr
\nu &= \pmns^\dagger\left( \ccal n_L - \scal N_L \right)\,,
\end{align}
where $\pmns$ is the usual PMNS matrix, and $ \scal $ and $ \ccal $ are diagonal $3\times 3 $ mixing matrices that obey
\beq
\scal^2 + \ccal^2 = \mati  \,.
\eeq
In terms of these quantities
\beq
\mf = \Lam \ccal\,, \quad Y\up\nu = \sqrt{2} \frac\Lam{\vh} \pmns^\dagger \scal\,,
\eeq
where $\vh$ denotes the Higgs \vev.

As a last simplification we will assume that the $z$ Yukawa couplings are real; in this case the model has 11 parameters: $\{\Lam,\,\mfe,\,\msc,\,\mu\}(4),~\{z,\,\scal,\,\lx\}(7) $.

The various interaction terms involving the $n_L$ and $N$ take the form
\bal
Z~\mbox{couplings:}& \quad
  - \frac g{2\cw} \left[ \bar n_L \ccal^2  \slashed Z n_L 
 + \bar N_L  \scal^2 \slashed Z N_L 
+ \left(\bar n_L \ccal \scal  \slashed Z N_L +{\rm H.c.} \right) \right]\,.
\mcr
 W~\mbox{couplings:}&\quad
  - \frac g{\sqrt{2}} \left[ \bar e \slashed W \pmnsd  \ccal n_L  - \bar e \slashed W \pmnsd \scal  N_L+ {\rm H.c.} \right] \,.
\mcr
\mbox{Yukawa~couplings:}& \quad
+ \frac\Lam \vh H \left[  \bar N_R  \scal \ccal n_L    -  \bar N_R  \scal^2  N_L   +  {\rm H.c} \right] \,.
\mcr
\mbox{DM~couplings:}& \quad
 + \left[\bar\Psi_\pm \Phi  z \scal n_L   +  \bar\Psi_\pm \Phi  z  ( \ccal   P_L  + P_R)N   +  {\rm H.c} \right] +\lx |\phi|^2 |\Phi|^2\,.
 \label{eq:couplings}
\end{align}

We identify the $ n_L$ with the observed neutrinos, however, these are massless, as noted above --  but this can be easily remedied by introducing a small Majorana mass term for the $ \fcal $: $\bar\fcal M_{\tt Maj} \fcal^c + \hc $ The effect is to slightly break the degeneracy of the $N$ and to give a Majorana mass to the $n_L$, whose form is the same as the one obtained in the inverse see-saw scheme~\cite{Wyler:1982dd,Mohapatra:1986bd,Ma:1987zm}. The $ M_{\tt Maj} $ term represent a soft and explicit breaking of lepton number, so the smallness of the neutrino masses is (technically) natural; since this mass matrix is arbitrary, it can be used to  generate the observed masses and mixing angles in the neutrino sector.

As a matter of notation we find it convenient to define
\beq
\r ij = \left( \frac {m_{\tt i}}{m_{\tt j}} \right)^2\,,
\label{eq:rij}
\eeq
so that $ \r NZ = (\Lam/\mz)^2$, etc.

\begin{figure}[ht]
$$ \includegraphics[scale=0.3]{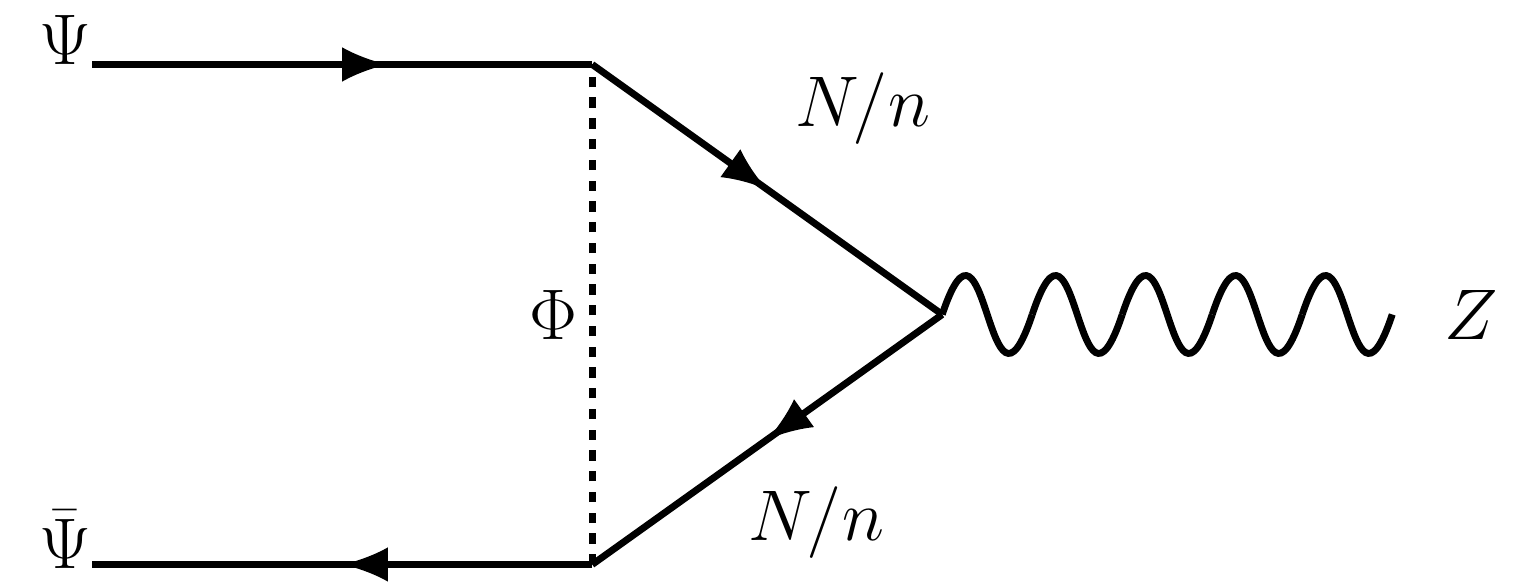}\quad \includegraphics[scale=0.3]{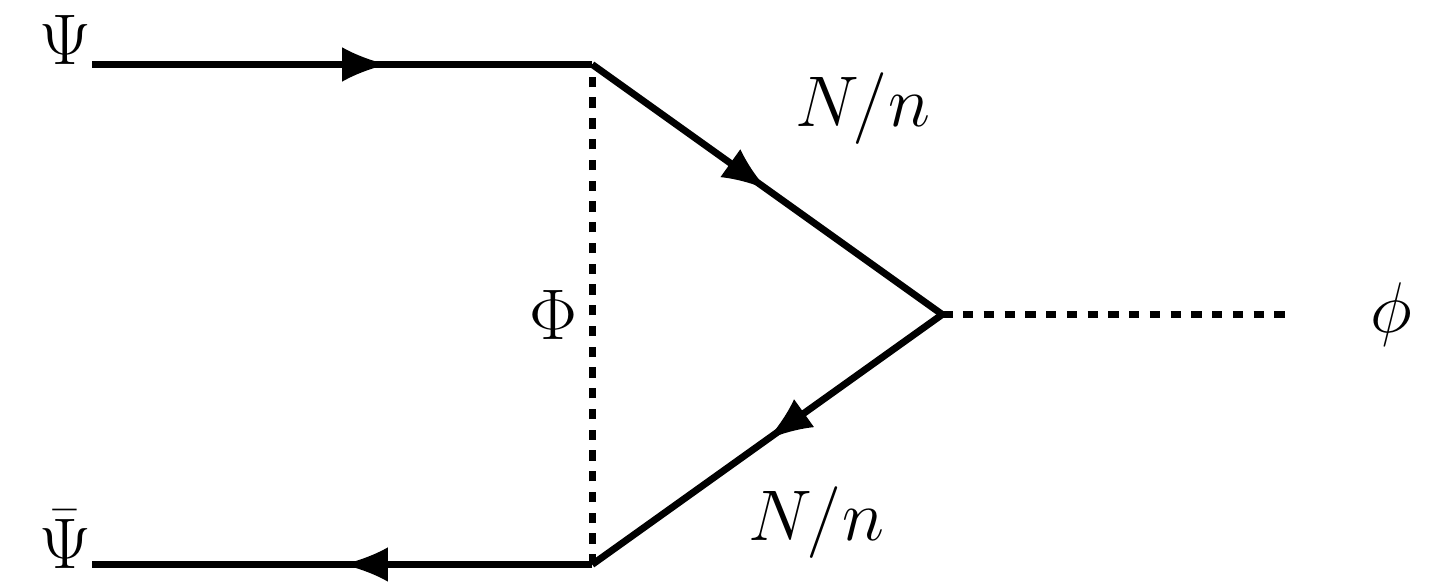}\quad \includegraphics[scale=0.3]{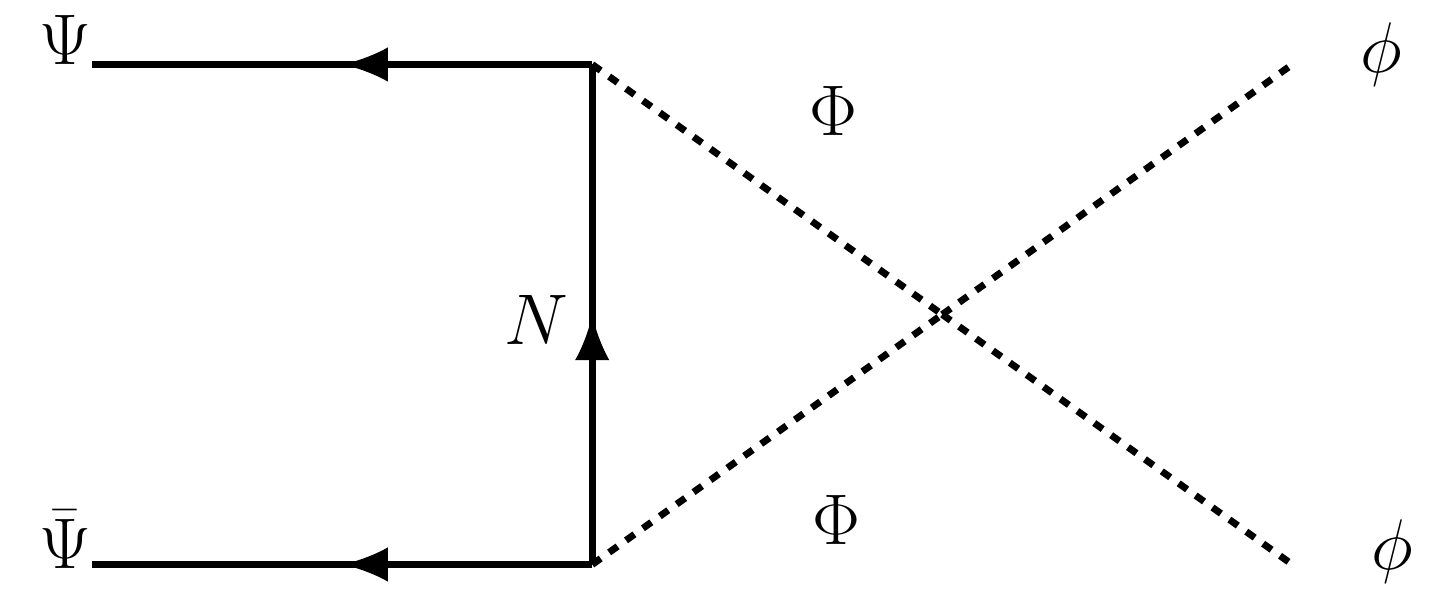}$$
\caption{Loop graphs generating the  $ \Psi \Psi Z$ and $ \Psi \Psi H $  couplings.}
\label{fig:z.h.couplings}
\end{figure}

\subsection{Loop-induced couplings}
\label{sec:loop.couplings}
The above model has no tree-level couplings of the  DM ($ \Psi $) to the $Z$ and $H$ bosons. These couplings are generated at 1 loop by the graphs in Fig. \ref{fig:z.h.couplings}. Assuming zero external momenta a straightforward calculation gives~\cite{Gonzalez-Macias:2016vxy}
\bal
\lcal_{\tt DM-Z} & =  - \frac g{2 \cw} \bar\Psi_\pm \slashed Z  \left( \epsilon_L   P_L+ \epsilon_R   P_R \right) \Psi_\pm \,;\mcr
\lcal_{\tt DM-H} & =  \epsilon_H \bar\Psi_\pm \Psi_\pm H\,,
\end{align}
where (see \cref{eq:rij}) 
\bal
\epsilon_R &= - \frac{ (z \scal^2  \ccal^2   z^T)}{32\pi^2}  \,  \, \frac{1-r_{\Phi {\tt N}}+\ln r_{\Phi {\tt N}}}{(1-r_{\Phi {\tt N}})^2}\,; \mcr
\epsilon_L &= \frac{(z    \scal^2    z^T)}{16\pi^2}  \, \frac{1-r_{\Phi {\tt N}}+r_{\Phi {\tt N}} \ln r_{\Phi {\tt N}}}{(1-r_{\Phi {\tt N}})^2} \,; \mcr
\epsilon_H &=  - \inv{8\pi^2} \frac\Lam\vh   \left\{    (z    \scal^2  \ccal    z^T)  \frac{1-r_{\Phi {\tt N}}+ r_{\Phi {\tt N}} \ln r_{\Phi {\tt N}}}{(r_{\Phi {\tt N}}-1)^2  }  + \half \lx \frac{\vh^2}{\Lam^2} (z    \ccal    z^T)  \frac{1 - r_{\Phi {\tt N}} + \ln r_{\Phi {\tt N}}}{(r_{\Phi {\tt N}}-1)^2} \right\} \,.
\end{align}

\subsection{DM self-interactions:}
\label{sec:DM-si}
 The strong-interactions of the SIDM paradigm are generated in this model by $ \Psi $ scattering mediated by $V$ exchange. There are two such reactions: $ \Psi _\pm \Psi _\pm \to \Psi_\pm \Psi_\pm $ and $ \Psi_+ \Psi_- \to \Psi_+ \Psi_- $, with cross section $ \sigma_{\tt r}$ and $ \sigma_{\tt a}$, respectively  (the first is the same  as M\"oller scattering with a massive photon). The calculation is straightforward, using 
\beq
\bP = \sqrt{1 - \frac{4 \mfe^2}s}\,,
\eeq
and neglecting the DM mass difference,  we find
\bal
\frac{\sigma_{\tt r}}\mfe&=
\frac{g^4}{4 \pi s \mfe } \Biggl\{ 
\frac{(2s + 3\mv^2) s \bP^2+ 2(\mv^2+2\mfe^2)^2}{2\mv^2(\mv^2+s \bP^2)} \mcr
&\hspace{1in}-\frac{  (s   \bP^2+ 2\mv^2) (3\mv^2+ 4\mfe^2) +2(\mv^2+2\mfe^2)^2- 4 \mfe^4  }{s\bP^2 \left(2 \mv^2 + s\bP^2\right)}\ln\left(1+\frac{s \bP^2}{\mv^2}\right)\Biggr\}\,;\mcr
\frac{\sigma_{\tt a}}\mfe&=\frac{g^4}{4 \pi  s  \mfe } \left\{ 
\frac{ (2s + 3\mv^2) s \bP^2 + 2(\mv^2+2\mfe^2)^2}{ 2\mv^2\left(\mv^2 +s \bP^2 \right)} -  \frac{ \left(\mv^2+s\right)}{s\bP^2} \ln\left(1+\frac{s \bP^2}{\mv^2}\right) \right\}\,.
\label{eq:sdm}
\end{align}
These cross sections are enhanced when $ \mfe \gg \mv $ and the relative velocity $ \bP$ is small; in this regime the $V$ interactions generate the required strong interactions.

Since $ \Psi_+$ and $ \Psi_- $ have but a small mass difference, and have identical couplings, they will have the same relic abundance density  $n$. In this case the effective DM-DM cross section will be $ (\sigma_{\tt r} + \sigma_{\tt a} )/2 $. To see this, note that a $ \Psi_+ $ moving with speed {\tt v}; in a time $ \delta t $ it will have $ n \sigma_{++} {\tt v} \delta t$ interactions with other $ \Psi_+$, and $ n \sigma_{+-} {\tt v} \delta t$ interactions with  the $ \Psi_-$; the total number of interactions will be then (using $ n= n_{\tt DM}/2 $),
\beq
\frac{n_{\tt DM} }2 \left(\sigma_{\tt r} +\sigma_{\tt a}\right)  {\tt v} \delta t = n_{\tt DM} \sigma_{\tt eff} {\tt v} \delta t \then \sigma_{\tt eff} = \frac{\sigma_{\tt r} +\sigma_{\tt a}}2\,.
\label{eq:seff}
\eeq
Note that $ \sigma_{\tt eff} $ depends on the relative velocity {\tt v}.

Existing data constraints the SIDM cross section for  galaxy clusters and for dwarf and low-surface-brightness galaxies; since the typical velocity in each environment is different, the cross section must have an appropriate velocity-dependence. The central values of the cross sections and velocities are ~\cite{Kaplinghat:2015aga} 
\beq
\left. \frac{\sigma_{\tt eff}}\mfe \right|_{\rm galaxy} \hspace{-.25in}=  1.9 \frac{\cm^2}{\gr} \,, ~ 
\left. \frac{\sigma_{\tt eff}}\mfe \right|_{\rm cluster}\hspace{-.25in} = 0.1 \frac{\cm^2}{\gr} \,;\quad
\left. \bP \right|_{\rm galaxy} =  3.3 \times 10^{-4} \,,~ 
\left. \bP \right|_{\rm cluster} = 5.4 \times 10^{-3}\,.
\eeq
Fitting \cref{eq:sdm,eq:seff} we find~\footnote{These relations imply $\mv g^{-4/3}= 0.144   \, \gev$, whose  significance is unclear.}
\beq
\mv = \frac\mfe{443}\,, \qquad g = \left( \frac\mfe{64 \, \gev} \right)^{3/4}\,.
\label{eq:mv.g.mfe}
\eeq
 These expressions have significant errors; using~\cite{Kaplinghat:2015aga} we estimate 
\beq
443 \to (116,\,1557) \,, \qquad 
64 \, \gev  \to (17,\,225)\, \gev\,.
\label{eq:mv.g.errors}
\eeq
In our numerical calculations we will be conservative and assume that these are uncertain by up to a factor of 3 (e.g., the first ranges from $443/3$ to $3*443$).

The DCC symmetry, despite being softly broken, is very effective in limiting the number of couplings of the $V$ that can have any phenomenological significance. For example, $V-Z$ and $ V-\gamma $ mixings occur only at 2 and 3 loops, respectively, and can be ignored. The only interesting 1-loop vertex is considered in the next section.

\subsection{Decay of the $V$}

In the absence of the DCC breaking term $ \propto \mu $ in \cref{eq:lag}, the massive dark photon $V$ is stable, which presents something of a problem: once it decouples from the $\Psi$, its abundance would be fixed., and since it is also light [cf. \cref{eq:mv.g.mfe}]~\footnote{We will see later that $ \mfe < O(10 \gev) $, whence $ \mv$ will be in the keV range.}, its presence would make the model inconsistent with big-bang nucleosynthesis (BBN)~\cite{Ahlgren:2013wba} and large scale structure formation (LSS)~\cite{Zhang:2015era} constraints. This is avoided when $ \mu \not =0$, that is, when the $ \Psi_\pm $ mass degeneracy is broken. 

\begin{figure}[ht]
$$ \includegraphics[scale=0.3]{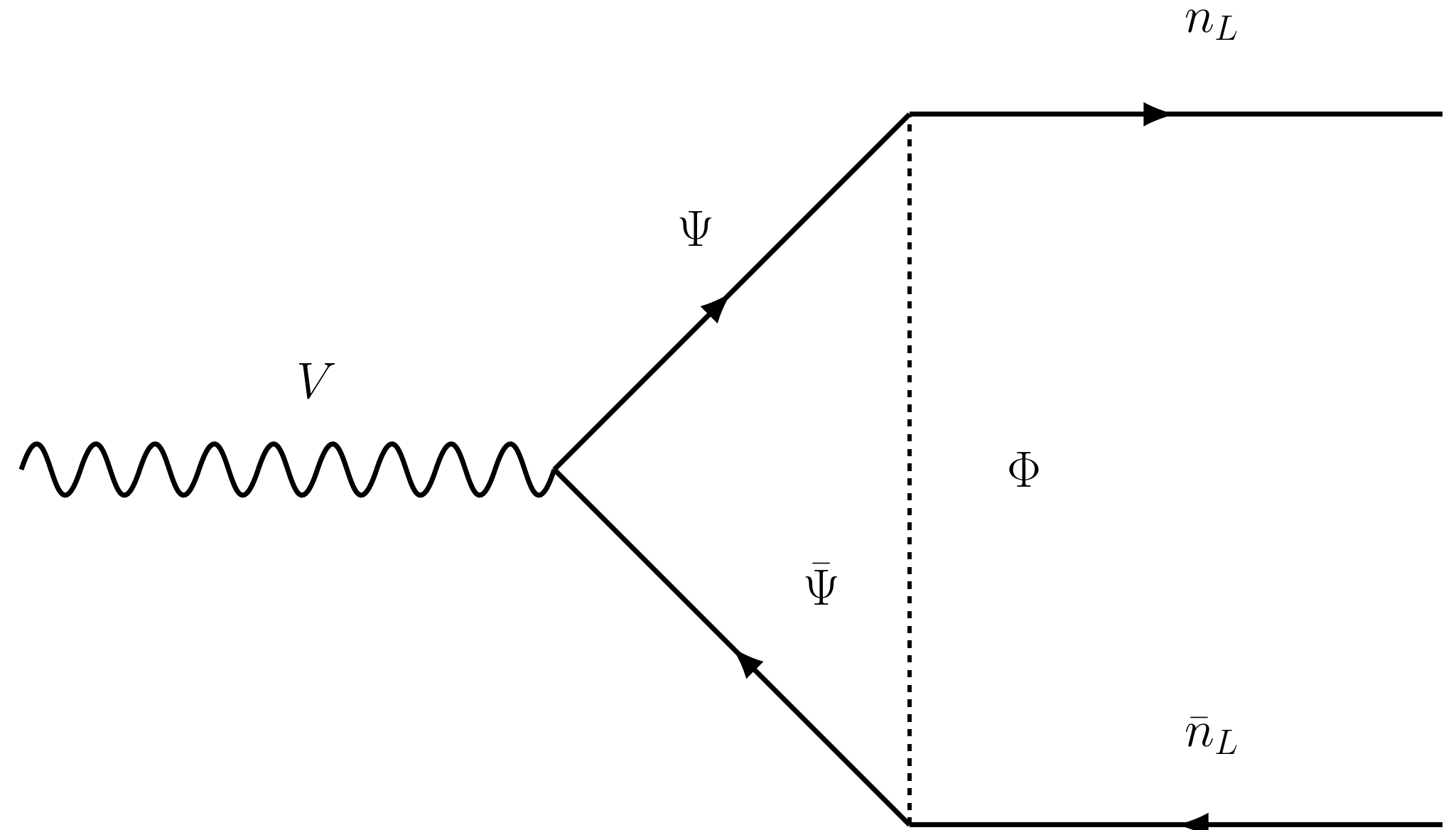} $$
\caption{Graphs responsible for a non-zero decay width for the $V$.}
\label{fig:v.width}
\end{figure}

In this case, the graphs in Fig. \ref{fig:v.width} give 
\beq 
\Gamma(V\rightarrow\bar{n}_{L}n_{L})=\frac{\mv}{6\pi}\left\{ \frac{g}{16\pi^{2}}\left[f\left(\frac{m_{+}}{\msc}\right)-f\left(\frac{m_{-}}{\msc}\right)\right]\right\} ^{2}\left(z \scal^{2}z^{\dagger}\right)^{2}\,,
\eeq 
where we assumed  $ \mfe \gg \mv,\,\mu $, and defined
\beq 
f(x)=\frac{1}{4}\left(\frac{x^{2}+1}{x^{2}-1}\right)-\left(\frac{x^{2}}{x^{2}-1}\right)^{2}\ln\,x\,.
\eeq 
As before, $m_{\pm}= \mfe  \pm \mu $ denote the mass of $\Psi_+$ and $\Psi_-$, respectively. The BBN and LSS constraints on this decay width are relatively mild:  $1/\Gamma( V \to \bar n_L n_L) < 1$s, which we adopt in the numerical calculations.

\subsection{Bound States}

The inclusion of a strong interaction between DM particles opens up the possibility that the $\Psi_+$ and  $\Psi_- $, having opposite charges, will form bound states. If this were to happen the strong interactions would be screened and the cusp problem would reappear. To avoid this we now consider the conditions for such bound states not to form.

In the non-relativistic limit, the $V$ exchange generates an attractive Yukawa potential between the $ \Psi_+$ and $ \Psi_- $:
\beq 
\mathcal{V}_{NR} = \frac{g^2}{4 \pi} \frac{e^{-\mv r}}{r} \,.
\eeq  
If a bound state is formed then its typical size is determined by the range of the potential, $ \sim 1/\mv $; it follows that the typical kinetic energy of the $ \Psi $ will be $ \sim \mv^2/\mfe $, while their potential energy would be $ \sim g^2 \mv/(4\pi) $. For the bound state to be unstable the kinetic energy must dominate: $g^2 \mv/(4\pi)  \lesssim \mv^2/\mfe $. These arguments are verified by exact calculations~\cite{An:2016gad,10.1093.ptep.ptx107} that give
\beq 
0.595 \frac{g^2}{4 \pi} < \frac{\mv}{\mfe}\,.
\eeq 
Using next the values of $ \mv $ and $g$ obtained in \cref{eq:mv.g.mfe} we find the following limit on $ \mfe$:
\beq 
\mfe < 8.4  \gev\,,
\eeq 
which is uncertain by up to a factor $ \sim 6 $.

Though $ \BS $ bound states are allowed for larger $ \mfe $, this does not necessarily imply that they will form. Formation occurs through the reactions $ \Psi_\pm\, \Psi_+\, \Psi_- \to \BS\, \Psi_\pm $, with a virtual $V$ exchange, or $ \Psi_+\, \Psi_- \to \BS + V $, with the (real or virtual) $V$ decaying subsequently to neutrinos. Calculating the rate for these reactions and determining the extent to which they affect the cusp problem in galactic DM distributions lies outside the scope of this paper. Here we will limit ourselves to the study of the model in the region  $ \mfe <  10\, \gev  $ where bound states do not occur, and  which is often outside the mass range considered in WIMP models (see, e.g. \cite{Baer:2014eja} and references therein). It is also worth noting that for these low masses the ``neutrino floor'' background in direct detection experiments rises by about 5 orders of magnitude (cf. Fig. \ref{fig:DD_limits}), and will will study to what extent this can conceal this model in this  region of parameter space.

\section{Electroweak constraints}
\label{sec:ew-constraints}
In this section we summarize the constraints derived from high precision data on the invisible decay fo the $Z$ and the Higgs, and from $W$-mediated meson decays; most of the results are the same as for an earlier simpler version of the model~\cite{Gonzalez-Macias:2016vxy}. These effects are produced by the mixing (upon \ssb) of the \sm\ neutrino field with the mediators $ \fcal $, which alters the couplings of the light mass eigenstates $n_L$ to the $W$ and $Z$, and introduces a coupling to the $H$ absent in the \sm.

\subsection{$Z$ invisible decay}
The addition of singlet Dirac fermions $N$ to the SM generate non-universal, though flavor diagonal, neutrino ($n$) couplings to the $Z$ proportional to $\ccal^2 $. In particular, the invisible $Z \to nn $ width will be proportional to tr$(\ccal^4) $.  The experimental value $\Gamma(Z\to {\rm inv})=499.0 \pm 1.5\, \mev$~\cite{Tanabashi:2018oca} for the invisible width of the $Z$ then generates a stringent bound on the parameters of the model when $ \mz < \Lam $; if the $Z$ decays involving the $N$  are kinematically allowed, the constraints are somewhat weaker.

Given a coupling of the form $ \bar\psi_1 \slashed Z ( a + b \gamma_5) \psi_2 $ we find that, if $ \mz > m_1 + m_2 $, 
\bal
\Gamma( Z \to \psi_1 \psi_2) = \frac{\left(|a|^2+|b|^2\right) \mz}{24\pi } \Biggl[ 2& -  \r1Z - \r2Z  -  \left(  \r1Z-\r2Z \right)^2 \mcr
& - 6 \frac{|a|^2 - |b|^2 }{|a|^2 + |b|^2} \sqrt{\r1Z\, \r2Z}\Biggr] \sqrt{ \lambda \left(1,\,\r1Z,\,\r2Z \right)}\,,
\end{align}
where $ \lambda(u,v,w) = u^2 + v^2 + w^2 - 2 uv - 2 vw - 2 wu $. For the case of degenerate $N$ this gives
\bal
&\Gamma( Z \to nn) = \Gamma_0 \tr{\ccal^4} \,;\qquad \Gamma_0 =  \left( \frac g{2\cw} \right)^2 \frac\mz{24\pi}\,, \mcr
&\Gamma( Z \to NN) = \Gamma_0 \tr{\scal^4} (1 -    \r NZ) \sqrt{ 1-4\r NZ}\, \theta(1-4\r NZ) \,,  \mcr
&\Gamma( Z \to Nn) = \Gamma_0 \tr{\ccal^2 \scal^2} \left( 2 +  \r NZ \right) (1-\r NZ)^2\, \theta (1 - \r NZ ) \,,
\end{align}
so that the change in the invisible decay width of the $Z$ is given by
\bal
\frac{\Gamma(Z\to{\rm inv})  }{\Gamma_{\tt SM}(Z\to{\rm inv}) } -1
=  \inv3 \Bigl[- \tr{\scal^2(\mati + \ccal^2)}& + \tr{\scal^4} (1 -    \r NZ) \sqrt{ 1-4\r NZ} \, \theta(1-4\r NZ) \mcr 
& + \tr{\ccal^2 \scal^2} \left( 2 + \r NZ  \right) (1-\r NZ)^2 \,\theta(1-\r NZ) \Bigr]\,;
\end{align}
 current experimental limits \cite{Tanabashi:2018oca} requires $ |\Gamma(Z\to{\rm inv})/\Gamma_{\tt SM}(Z\to{\rm inv}) -1|< 0.0093$.

\subsection{$H$ invisible decays}
A general coupling of the form $ \bar\psi_1  (a + b \gamma_5) \psi_2 H  $ gives
\beq
\Gamma( H \to \psi_1 \psi_2)  = \frac{\left( |a|^2 + |b|^2 \right)\mh}{8\pi  } \left[ 1 - r_{1{\tt H}}  - r_{2{\tt H}} - 2  \frac{ |a|^2 - |b|^2 }{ |a|^2 + |b|^2  } \sqrt{r_{1{\tt H}}  r_{2{\tt H}}} \right] \sqrt{ \lambda( 1,r_{1{\tt H}}  , r_{2{\tt H}}) } \,.
\eeq
Using this and \cref{eq:couplings} we obtain 
\bal
& \Gamma(H\to \Psi\bar\Psi)= \frac{\mh\epsilon_H^2}{8\pi} (1 - 4 \r\Psi H )^{3/2} \theta (1 - 4\r\Psi H) \,, \mcr
& \Gamma(H \to n,N )  = \frac{\mh^3}{4\pi \vh^2} \left[ \r NH (1-\r NH ) \tr{ \scal^2 \ccal^2} \theta(1-\r NH ) \  + \half (1 - 4\r NH )^{3/2} \tr{\scal^4} \theta(1-4\r NH ) \right] \,,  \mcr
& \Gamma(H \to \Phi \Phi)  = \frac{(\vh \,  \lx)^2 }{16\pi \mh} \sqrt{1 - 4 \r\Phi H }\, \theta( 1-4 \r\Phi H )\,,.
\label{eq:h.inv}
\end{align}
The first width in \cref{eq:h.inv}  is  negligible because of the $ \epsilon_H^2$ prefactor.

 The total  width of the $H$ is then $ \Gamma(H) = \Gamma(H)_{\tt SM} + \Gamma(H \to n,N ) + \Gamma(H \to \Phi \Phi)$, with the  SM contribution equal to $ 4\, \mev $~\cite{Tanabashi:2018oca}; given that the limit on the invisible branching ratio is $24\%$, we find $ \Gamma(H \to n,N ) + \Gamma(H \to \Phi \Phi)<1.26 ~\mev $. Then, for degenerate $N$,

\bal
4.89 \times 10^{-4} >   \Bigl| \r NH  (1-\r NH  ) \tr{ \scal^2 \ccal^2} \theta(1-\r NH  )  +&  \half (1 - 4\r NH  )^{3/2} \tr{\scal^4} \theta(1-4\r NH  ) \mcr
& \quad + 1.93 \lx^2 \sqrt{1 - 4\r\Phi H  } \theta( 1-4\r\Phi H   ) \Bigr| \,.
\end{align}

\subsection{$W$-mediated decays}
The second line in \cref{eq:couplings} shows that charged current interactions of the leptons and the $W$ boson are also modified: using $ r,s$ as flavor indices, the vertex involving a charged lepton $e_{L\,r}$ and a neutrino mass eigenstate $n_{L\,s}$ contains a factor $(\pmnsd\ccal)_{rs}$. This then implies (we assume that $ \Lam > m_\tau $)
\beq
\Gamma(\ell_r \to \ell_s \bar\nu \nu)  \simeq (1 - \Delta_r - \Delta_s) \Gamma_{\tt SM}(\ell_r \to \ell_s \bar\nu \nu)  \,; \quad \Delta_r =  \left( \pmnsd \scal^2\pmns  \right)_{rr} > 0\,,
\label{eq:w.med.dec}
\eeq
(no sum over $r$ in the last expression). Note that the assumption $ \Lam > m_\tau $ precludes the possibility of there being cancellations between the $n$ and $N$ contributions to these decays.

\begin{figure}[ht]
$$\includegraphics[width=2.3in]{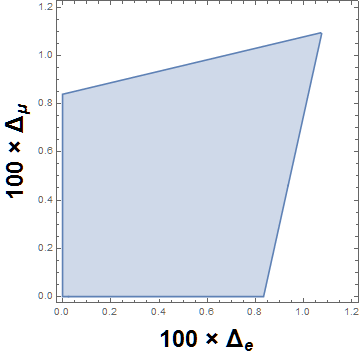}$$
\caption{Limits on $ \Delta_r$ [cf. \cref{eq:w.med.dec}] derived from $W$-mediated decays.}
\label{fig:w.constraints}
\end{figure}

We define $ R_{u \to X} = B(u \to X)/B_{\tt SM}(u \to X) -1 $; then, for the specific decays of interest, we find (to 3$\sigma$),
\bal
R_{\tau \to \mu\nu\bar\nu}  \simeq B_{\tt SM}(\tau \to e\nu\bar\nu) \Delta_e -  \left[1-B_{\tt SM}(\tau \to \mu\nu\bar\nu)\right]\Delta_\mu  &\then  | 0.8223\,\Delta_\mu - 0.1958\,\Delta_e | \le 0.0069\,,\mcr
R_{\tau \to  e\nu\bar\nu}  \simeq   B_{\tt SM}(\tau \to \mu\nu\bar\nu) \Delta_\mu- \left[1-B_{\tt SM}(\tau \to e\nu\bar\nu)\right] \Delta_e &\then | 0.1777\,\Delta_\mu - 0.8042\,\Delta_e | \le 0.0067\,,\mcr
R_{\pi\to \mu\nu} \simeq  B_{\tt SM}(\pi\to e\nu)(\Delta_\mu - \Delta_e) &\then  |\Delta_\mu - \Delta_e| \le 0.010\,.
\end{align}
These constraints are summarized in Fig. \ref{fig:w.constraints}. We note that the limit derived from $ \pi \to \mu e $ is not competitive: $|\Delta_\mu - \Delta_e| \le 48.8 $. Also, though the  uncertainty in $ \Gamma(\mu \to e\nu\bar\nu)$ is very small,  it does not lead to a constraint on $ \Delta_e + \Delta_\mu$, since this decay is used as input data to fix the value of $G_{\tt F}$. One could use collider measurements of $ \mw $ and $g_2$ (the $\su2_L$ coupling constant in the SM) to predict this width, but the uncertainty is much larger and the limits are again not competitive.

\subsection{Muon anomalous magnetic moment.}
The new $NNW$ vertices, and the $ \ccal$ factors for the $nnW$ vertices in \cref{eq:couplings} generate  contribution to the anomalous magnetic moment of the muon, $ a_\mu $. Using the results of ~\cite{Leveille:1977rc} it is straightforward to see that
\beq
\Delta a_\mu = \frac{ G_{\tt F}\, m_\mu^2}{\sqrt{2} \, 8\pi^2} \Delta_\mu \left[ F(r_{\tt N W}) - F(0) \right]\,,
\eeq
where $ \Delta_\mu = \Delta_{r=2} $ is defined in \cref{eq:w.med.dec} and
\bal
F(w) &= \int_0^1 dx \frac{2 x^2(1+x) + x(1-x)(2-x) w- x^2(x-1)k}{k x^2 +(1-k) x+ (1-x) w}\,; \quad k = \left( \frac{m_\mu}\mw \right)^2  \,, \mcr
&\simeq \int_0^1 dx \frac{2 x^2(1+x) + x(1-x)(2-x) w}{ x+ (1-x) w}\,, 
\end{align}
so that
\beq
F(w) - F(0) \simeq \frac{10 - 33 w + 45 w^2 - 4 w^3}{6(1-w)^3} + \frac{3 w^3 \, \ln w}{(1-w)^4} - \frac53\,,
\eeq
and this ranges from $0$ when $w=0$ to $-1$ when $ w \to\infty $. Then
\beq
| \Delta a_\mu| \le \frac{ G_{\tt F}\, m_\mu^2}{\sqrt{2} \, 8\pi^2} \Delta_\mu = 1.17 \times 10^{-9} \Delta_\mu\,.
\eeq
The constraints derived form $W$-mediated decays require $ \Delta_\mu \lesssim 10^{-2}$ (see Fig. \ref{fig:w.constraints}) so $| \Delta a_\mu|  \lesssim 10^{-11} $, while the current error~\cite{Tanabashi:2018oca} is $ (\pm 5.4 \pm 3.3) \times 10^{-10} $. The anomalous magnetic moment limits do not produce a competitive bound now, but may do so with the upgraded Fermilab experiment~\cite{Grange:2015fou}\footnote{It does not  explain either the new anomaly in the  magnetic moment of the electron~\cite{Parker:2018vye} since that will be suppressed by a factor $(m_e/m_\mu)^2$ with respect to the $(g-2)_\mu$.}.

\section{Relic abundance.}
\label{sec:rel-ab}
As the universe expands there will come a time when the $ \Psi_\pm $ will cease to be in chemical equilibrium with the SM or with the dark photon sea. Still, we expect the  interactions between $\Psi_+$ and $\Psi_-$ will keep them in equilibrium with each other and, since they have (approximately) equal mass and couplings with $V$, they will have the same relic abundance; in the following we denote by $ n_\Psi $ the {\em total} DM number density, adding the contributions from the $ \Psi_+$ and $ \Psi_-$.

\begin{figure}
\begin{centering}
\includegraphics[scale=0.3]{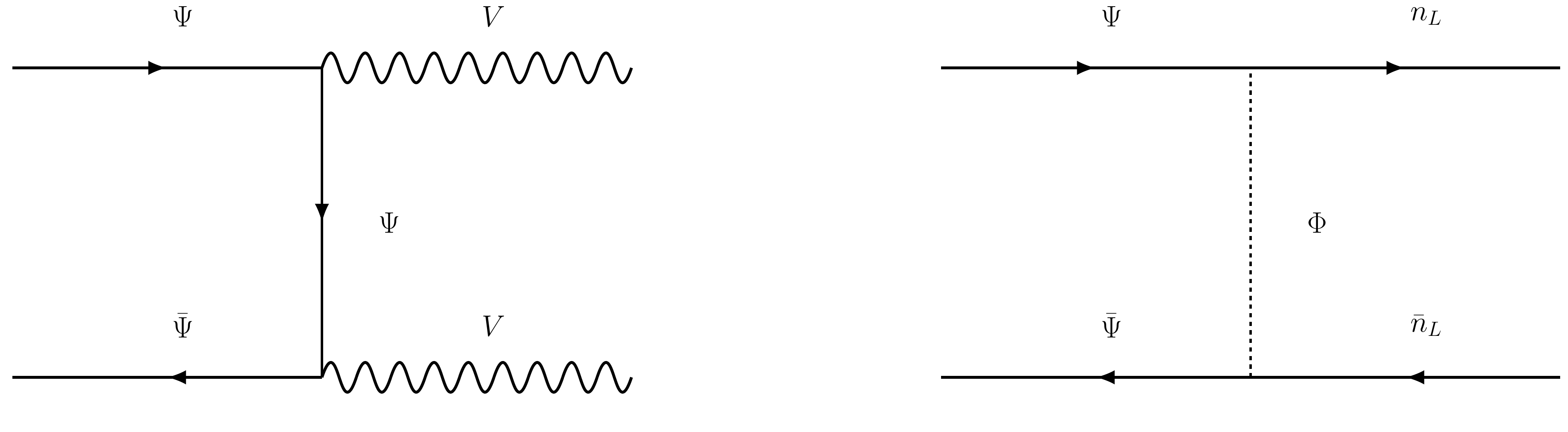} 
\par\end{centering}
\caption{Diagrams giving the leading contributions to the relic abundance cross sections.}
\label{fig:RA.graphs}
\end{figure}

The processes that determine the relic abundance are (see Fig. \ref{fig:RA.graphs})  then $ \Psi \bar\Psi \to n_L \bar n_L $  and $ \Psi \bar\Psi \to VV$ , for which the cross sections are
\bal
\sigma_{\Psi\Psi\to nn} &=\frac{\left( z\scal^2 z^{T} \right)^2}{64\pi s\bP}\left[\frac{1+2y(1+y)-\bP^{2}}{(1+y)^{2}-\bP^{2}}+\frac{y}{\bP}\ln\left(\frac{1-\bP+y}{1+\bP-y}\right)\right] \,, \mcr
\sigma_{\Psi \Psi \to {\tt VV}} 
&= \frac{g^4}{8\pi s} \frac\bv\bP \left[
\frac{ s \mfe^2 +4(\mv^4-2 \mv^2 \mfe^2- 2 \mfe^4)}{s \mfe^2 + \mv^2(\mv^2 - 4 \mfe^2)} + \frac{4( \mv^2 + \mfe^2)}{s \bv \bP} \ln \left| \frac{1 + \bv^2 + 2 \bP \bv}{1 + \bv^2 - 2 \bP \bv} \right| \right]  \,,
\end{align} 
where 
\beq 
y=\frac{2(\msc^2-\mfe^2)}{s}\,; \qquad \bP=\sqrt{1-\frac{4 \mfe^2}{s}}\,; \qquad \bv=\sqrt{1-\frac{4 \mv^2}{s}}\,.
\eeq

Since we are considering DM masses smaller than those for the $Z$ and $H$, there will be no resonant contributions to the relic abundance calculations, and the usual approximations \cite{Kolb:1990vq} can be reliably used. After a straightforward calculation we find
\bal
\vevof{{\rm v} \,\sigma_{\Psi \Psi \to nn}} &\simeq  \frac{\left(z \scal^2 z^T \right)^2}{32 \pi (\r\Psi\Phi + 1)^2 \mfe^2}  \,, \mcr
\vevof{{\rm v} \, \sigma_{\Psi \Psi \to {\tt VV}}} &\simeq  \frac{g^4}{16\pi \mfe^2 }\,,
\end{align}
where we summed over all final neutrino states and took $ \mfe \gg \mv $ (cf. \cref{eq:mv.g.mfe}); since there is no temperature dependence to lowest order, these are s-wave reactions.

We follow the usual prescription for abundance calculation via the Boltzmann Equation:
\beq
\frac{dn_{\Psi}}{dt}+3Hn_{\Psi}=- \sigma_0\left[ n_\Psi^2-\left(n_\Psi\up{\tt eq}\right)^{2}\right]\,,
\eeq
where
\beq
\sigma_0 = \half \vevof{{\rm v} \, \sigma_{\Psi \Psi \to nn}}+\inv4 \vevof{{\rm v} \, \sigma_{\Psi \Psi \to {\tt VV}}} = 
 \frac{g^4 + \left[ z \scal^2 z^T /(\r\Psi\Phi + 1)\right]^2}{64 \pi  \mfe^2}\,.
 \label{eq:sig0}
\eeq 
Using the standard freeze-out approximation ~\cite{Kolb:1990vq}, the relic abundance $\Omega_\Psi $ is given by:
\beq
\Omega_\Psi h^2 = \frac{1.07 \times 10^9}\gev  \frac{ x_f  }{{\tt g}_{\star \tt s}\xi} \,; \quad \xi = \frac{ M_{\rm Pl} \sigma_0 }{\sqrt{{\tt g}_\star}}
\label{eq:Omapprox}
\eeq
where $M_{\rm Pl}$ denotes the Planck mass, $ {\tt g}_{\star \tt s} ,\, {\tt g}_\star $ denote, respectively, the relativistic degrees of freedom associated with the entropy and energy density~\footnote{For our numerical calculations we use the expression of ${\tt g}_\star $ in \cite{Gondolo:1990dk}, not the one from \cite{Kolb:1990vq}.} (for our case they are the same), and
\beq
x_f = \frac\mfe{T_f} = \ln \left(0.076 \mfe \xi \right) - \half \ln \left[ \ln \left(0.076 \mfe \xi \right) \right]\,,
\label{eq:xf}
\eeq
with $T_f$ the freeze-out temperature.
This expression for $ \Omega_\Psi $ can now be compared to the result inferred from CMB data obtained by the Planck experiment~\cite{Aghanim:2018eyx,Ade:2015xua}:
\beq
\Omega_{\rm Planck} h^2 = 0.12\pm0.003 \quad (3\sigma).
\label{eq:planck}
\eeq
Note, in particular, that a sufficiently large value of the dark-photon coupling $g$  will lead to DM under-abundance.
 
\section{Direct Detection}
\label{sec:dd}
In the model under consideration the DM-nucleon  scattering cross section responsible for a direct detection signal is generated by  (t-channel) $Z$ and $H$ exchanges associated with the loop-induced couplings listed in Sect. \ref{sec:loop.couplings}. Since the momentum transfer is much smaller than $ \mz $ and $\mh $ we can approximate the relevant interaction by
\beq 
\lcal_{\tt nucleon-DM} = \sqrt{2} G_{\tt F} \, \left[ \bar\Psi \gamma_\mu \left( \epsilon_L  P_L+ \epsilon_R P_R \right) \Psi \right] \, \left( \bar\prot \jcal^\mu_\prot \prot + \bar\neut \jcal^\mu_\neut \neut \right) + G_{\tt H} \bar\Psi  \Psi \left( \bar\prot  \prot + \bar\neut   \neut \right)\,,
\eeq
where $ \prot,\,\neut$ denote, respectively, the proton and neutron fields and~\footnote{In the expressions for $ \jcal^\mu_{\prot,\neut} $ we did not include a term $ \propto \Delta s \gamma^\mu \gamma_5$ since the current experimental values for $ \Delta s $ \cite{Airapetian:2006vy,Ageev:2007du,Maas:2017snj} are consistent with zero.}  \cite{Engel:1991wq}
\bal
\jcal^\mu_\prot &= \half \left[ 
\left(1 - 4 \sw^2 \right)  \gamma^\mu   +  g_A \left( \gamma^\mu - \frac{2 \mnu q^\mu}{m_\pi^2 + \qq^2} \right)\gamma_5 \right] \,, \mcr
\jcal^\mu_\neut  &=- \half \left[   \gamma^\mu  + g_A \left( \gamma^\mu - \frac{2 \mnu q^\mu}{m_\pi^2 + \qq^2} \right) \gamma_5   \right]\,,
\end{align}
with $ \mnu,\,m_\pi $ the nucleon and pion masses, $\qq$  the momentum transfer, $ g_A \simeq  -1.2723 $ the axial nucleon coupling \cite{Tanabashi:2018oca}, and \cite{Cheng:2014opa}
\beq
G_{\tt H} = - \frac{0.011 \epsilon_H}{\mh^2}\,.
\eeq
All isospin breaking effects in the Higgs-mediated interactions were ignored.

In the non-relativistic limit this becomes
\bal
\inv{\mfe \mnu}\left. \lcal_{\tt nucleon-DM} \right|_{\tt NR} =& 4 G_H   \mati_\Psi \,\mati_\nucl
+ \sqrt{2} \, G_{\tt F} (\epsilon_R + \epsilon_L)   \Biggl\{  \left[ - 2 \sw^2 +(1- 2 \sw^2) ] \tau_3 \right] \mati_\Psi \,\mati_\nucl \mcr
&\qquad\qquad  +  \tau_3 \left[ \ss_\Psi . \ss_\nucl - 4 \frac{(\qq.\ss_\Psi)(\qq.\ss_\nucl)}{m_\pi^2 + \qq^2} \right] \left( \frac{\epsilon_R - \epsilon_L}{\epsilon_R + \epsilon_L}  \right)g_A \Biggr\}\,,
\end{align}
where $ \tau_3 \to 1 $ for $ \prot$ and $ \tau_3 \to -1$ for $\neut$, $ \ss_{\Psi,\, \nucl}$ denote the spin operators for the DM and the nucleons. Using the notation and procedure described in \cite{Fitzpatrick:2012ix,Anand:2014kea} (see also \cite{Walecka:1975}) we find that the DM-nucleus cross section, which we denote by $ \sigma_\atm $ is given by
\bal
\sigma_\atm&= \frac{( \ma/\mnu)^2}{16 \pi (\ma + \mfe)^2}\biggl\{  \kappa^2 \left[ (1+b)^2 F_M\up{\prot,\prot} + (1-b)^2 F_M\up{\neut,\neut} + 2(1-b^2) F_M\up{\prot,\neut} \right] \mcr
& \quad + \frac{\kBB^2(Q^2-2Q+3)}{12} \left[ F_{\Sigma''}\up{\prot,\prot} +  F_{\Sigma''}\up{\neut,\neut} -2 F_{\Sigma''}\up{\prot,\neut} + 2 \left( F_{\Sigma'}\up{\prot,\prot} +  F_{\Sigma'}\up{\neut,\neut} -2 F_{\Sigma'}\up{\prot,\neut}\right) \right] \biggr\}\,,
\end{align}
where $\atm$ is the atomic number, $\ma \simeq \atm \mnu $ the nuclear mass, and
\bal
&\kappa = \sqrt{2}  G_{\tt F} \mfe \mnu \left[ 2 (\epsilon_L + \epsilon_R) \sw^2 - 2 \sqrt{2} \, \frac{G_{\tt H}}{G_{\tt F}} \right]\,, \quad Q = \frac{4 |\qq|^2}{|\qq|^2 + m_\pi^2}  \,, \mcr
& \kBB=\frac{G_{\tt F} (\epsilon_R - \epsilon_L) \mfe \mnu}{\sqrt{2}} g_A\,, \quad b = \frac{1 - 2\sw^2}{\sqrt{8}G_{\tt H}/[(\epsilon_L + \epsilon_R)G_{\tt F}] - 2 \sw^2}\,.
\end{align}
The DM-nucleon cross section is then defined \cite{Fitzpatrick:2012ix,Lewin:1995rx} as
\beq
\sigma_N = \left( \frac\mnu{\ma} \right)^2 \left( \frac{\mfe + \ma}{\mfe + \mnu} \right)^2 \inv{\atm^2} \sigma_\atm\,.
\eeq

If there are several isotopes, labeled by $I$, with abundances $ \alpha_I $, then $ F_X\up{N,N'} \to\,   ^I\hspace{-2pt}F_X\up{N,N'} $ and
\beq
\inv{\atm^2} F_X\up{N,N'} \to \sum_I \frac{\alpha_I}{\atm_I^2} \;   ^I\hspace{-2pt}F_X\up{N,N'} = f_X\up{N,N'}\,;
\eeq
so, defining
\bal 
f_1 =& f_M\up{\prot,\prot} + f_M\up{\neut,\neut} + 2f_M\up{\prot,\neut}\,,\mcr
f_2 =& f_M\up{\prot,\prot} - f_M\up{\neut,\neut}\,,\mcr
f_3 =& f_M\up{\prot,\prot} + f_M\up{\neut,\neut} - 2f_M\up{\prot,\neut}\,,\mcr
f_4 =& \left( f_{\Sigma''}\up{\prot,\prot} +  f_{\Sigma''}\up{\neut,\neut} -2 f_{\Sigma''}\up{\prot,\neut} \right) +  \left( f_{\Sigma'}\up{\prot,\prot} +  f_{\Sigma'}\up{\neut,\neut} -2 f_{\Sigma'}\up{\prot,\neut}\right) \,,
\end{align}
the expression for the DM-nucleon cross section takes the relatively simple form
\beq
\sigma_\nucl = \inv{16\pi^2(\mnu + \mfe)^2} \left[  \left( f_1 + 2 b f_2 + b^2 f_3 \right)\kappa^2 +  \frac{\kBB^2(Q^2-2Q+3)}{12}f_4 \right]\,.
\eeq
It is worth noting that the term $ \propto \kappa^2 $ is the spin-independent contribution, while that $ \propto \kBB^2 $ is the spin-dependent one. The expected suppression of the latter with respect to the former follows from $ f_4 \ll f_1 $. In the calculations we use the expressions for the $^I\hspace{-2pt}F_X\up{N,N'} $ provided in \cite{Fitzpatrick:2012ix} for Xe and Ge, and in \cite{Catena:2015uha}~\footnote{Note that there is a normalization factor of $ \pi $ difference between the conventions of \cite{Fitzpatrick:2012ix} and \cite{Catena:2015uha}; for example $F\up{\prot,\prot}_M$ in \cite{Fitzpatrick:2012ix} equals $ \pi \times \left[ W\up{0,0}_M + 2 W\up{0,1}_M  +  W\up{1,1}_M  \right] $ in \cite{Catena:2015uha}.} for CaWO$_4$:
{\small
\beq
\begin{array}{c||c|c|c|c}
\text{element} & f_1 & f_2 & f_3 & f_4 \times 10^4 \cr
\hline
\text{Xe}     & 0.995256  - 6.98794  \qq^2 & -0.177925 + 1.39348  \qq^2 & 0.031717  - 0.314739 \qq^2 & 0.142261 - 1.22925\qq^2 \cr
\text{Ge}     & 0.990137  - 6.97097  \qq^2 & -0.124142 + 0.960981 \qq^2 & 0.0161359 - 0.115939 \qq^2 & 0.156404 - 1.61629 \qq^2 \cr
\text{CaWO}_4 & 0.0624983 - 0.447775 \qq^2 & 0                          & 0                          & 0
\end{array}
\eeq
}
and we took $|\qq| = \mfe \times 300$km/s.

We note that the dependence of $ \sigma_\nucl$ on $ \mfe $ is simple and contained in the factor $ [\mfe/(\mfe+ \mnu)]^2 $, it also has a more complicated dependence on $ \msc,\,\Lam$ through the parameters $ \epsilon_{L,\,R,\,H}$.

In the numerical results below we used the experimental constraints on the direct detection cross section published by Xenon1T \cite{Aprile:2019dbj}, PandaX \cite{Cui:2017nnn}, CDMS \cite{Agnese:2013rvf} and CRESST \cite{Abdelhameed:2019hmk} for the range $ 0.36 \, \gev \le \mfe \le 10\,\gev $; in cases where the mass ranges of two experiments overlap we take the strictest limit. Specifically, we used:
\beq
\begin{array}{|l|rl|}
\text{Experiment} & \multicolumn{2}{|c|}{\mfe\text{ range (GeV)}}\cr
\hline
\text{Xenon1T} & \quad(6.06,&10.0) \cr
\text{PandaX}  & (4.12,&6.06) \cr
\text{CDMS}    & (1.61,&4.12) \cr
\text{CRESST}  & (0.36,&1.61) 
\end{array}
\label{eq:ranges}
\eeq
as illustrated in Fig. \ref{fig:DD_limits}.

\section{Numerical Results}
\label{sec:numerics}
The model being considered has in total 14 free parameters: $\mfe$, $\mu $, $\msc$, $ \Lam$, $\lx$, $\scal $ and $z$ (we assumed $\mv$ and $g$ are fixed by \cref{eq:mv.g.mfe,eq:mv.g.errors}). We will for simplicity assume that $z$ is real since all the observables we  consider depend only on the magnitudes $ z_i $, this reduces the number of parameters to 10. In this section we consider the region in parameter space  
\bal
&0.5 \gev \le \mfe \le 10 \gev \,, \qquad \mu = \frac\mfe{20}\,,\mcr
& \text{min}\{1.1 \mfe,\,\mfe + 2 \gev\} \le \msc < 500 \gev\,, \mcr 
&\text{min}\{1.1 \mfe,\,\mfe + 2 \gev\}  \le \Lam \le 1.5 \tev\,, \mcr
& |\lx| \le \pi\,, \quad
|S_i| < 1\,,\quad
|z_i|^2 \le 10 \quad (i=1,2,3);
\label{eq:region}
\end {align}
and determine the sub-region allowed by the various constraints listed above. This is frequently carried out by reducing the number of free parameters (e.g. fixing the $ \scal $ and taking all the $z_i$ equal \cite{Gonzalez-Macias:2016vxy}) and then doing a uniform scan in the reduced space. Here we follow a different route: we do not adopt any simplifying relations between the parameters (except $ \mu $), and concentrate on finding the boundary of the allowed sub-region; this then becomes a non-linear optimization problem that can be treated using standard techniques \cite{Pierre:1986}. In our calculations we use a publicly-available non-linear programming package NLOPT \cite{Johnson:NLOPT}.

\begin{figure}[ht]
$$ 
\begin{array}{ccc}
\includegraphics[scale=0.6]{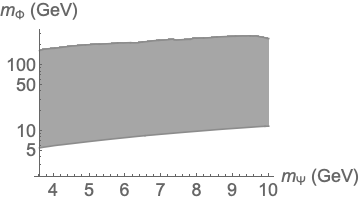}  & \qquad& \includegraphics[scale=0.6]{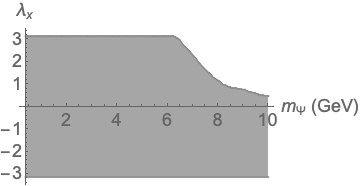}\cr
(a)&& (b) \vspace{20pt}\cr
\includegraphics[scale=0.6]{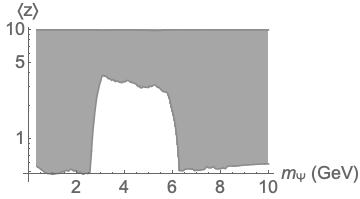}  & \qquad& \includegraphics[scale=0.6]{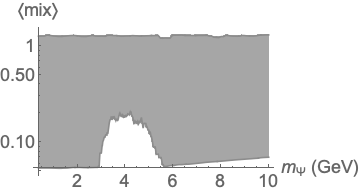}\cr
(c)&& (d) 
\end{array}$$
\caption{Projections of the allowed parameter region, $(a)$ in the $ \mfe-\msc$ plane, $(b)$  the $ \lx-\msc$ plane, $(c)$ the $\mfe-\vevof z $ plane (where $ \vevof z = |z|/\sqrt{3} $), and $(d)$  the $\mfe -  \vevof{\text{mix}} $ plane, where $\vevof{\text{mix}}$ is defined in \cref{eq:mix}. The unevenness in the curves are due to numerical inaccuracies.}
\label{fig:MPhi-mix}
\end{figure}

We define
\beq 
\vevof{\text{mix} }=\sum |z_i|^2 \scal_i^2\,, \quad \vevof z = |z|/\sqrt{3}\,,
\label{eq:mix}
\eeq
as measures of the mixing strength and Yukawa coupling of the mediators, and then obtain the projections of the allowed sub-region in the $ \mfe-\msc$, $ \mfe-\lx$,  $ \mfe- \vevof z $ and   the $ \mfe - \vevof{\text{mix}} $ planes. The results are presented in Figs. \ref{fig:MPhi-mix} $(a) -(d)$ respectively.  In the $ \mfe-\Lam$ plane the constraints allow the full area indicated in \cref{eq:region}; that is, for each point in this area there are values of the other parameters for which all constraints are satisfied (in general these values change for each choice of $\mfe $ and $ \Lam $). The features in figures $(b),\,(c),\,(d)$ at  $ \mfe\sim  4\,\gev$ and $ \mfe\sim 6\,\gev $ are due to the changes in the constraints of the direct-detection cross section (cf. \cref{eq:ranges}).

\begin{figure}[ht]
$$ 
\includegraphics[scale=0.4]{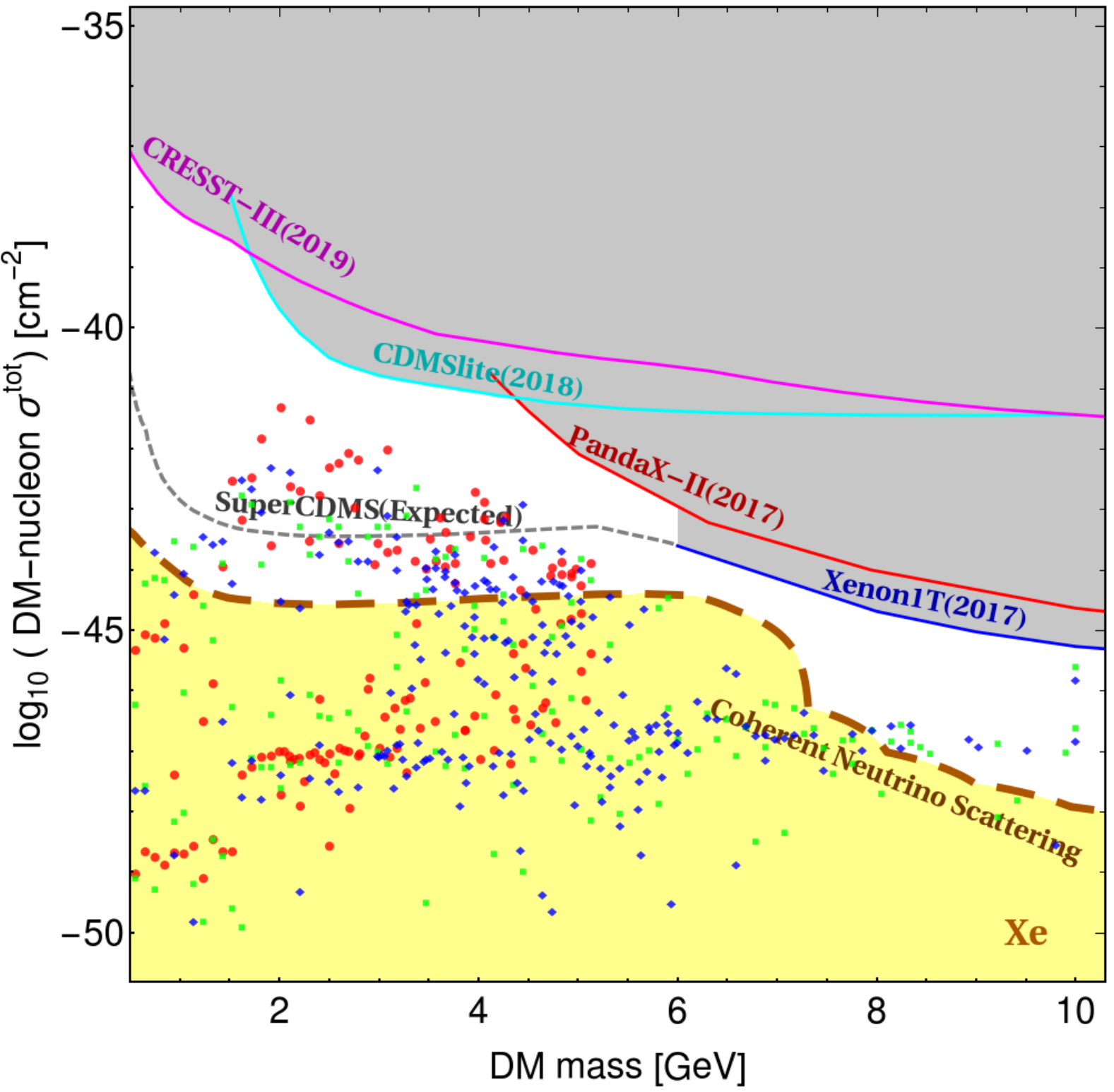} \qquad
\includegraphics[scale=0.4]{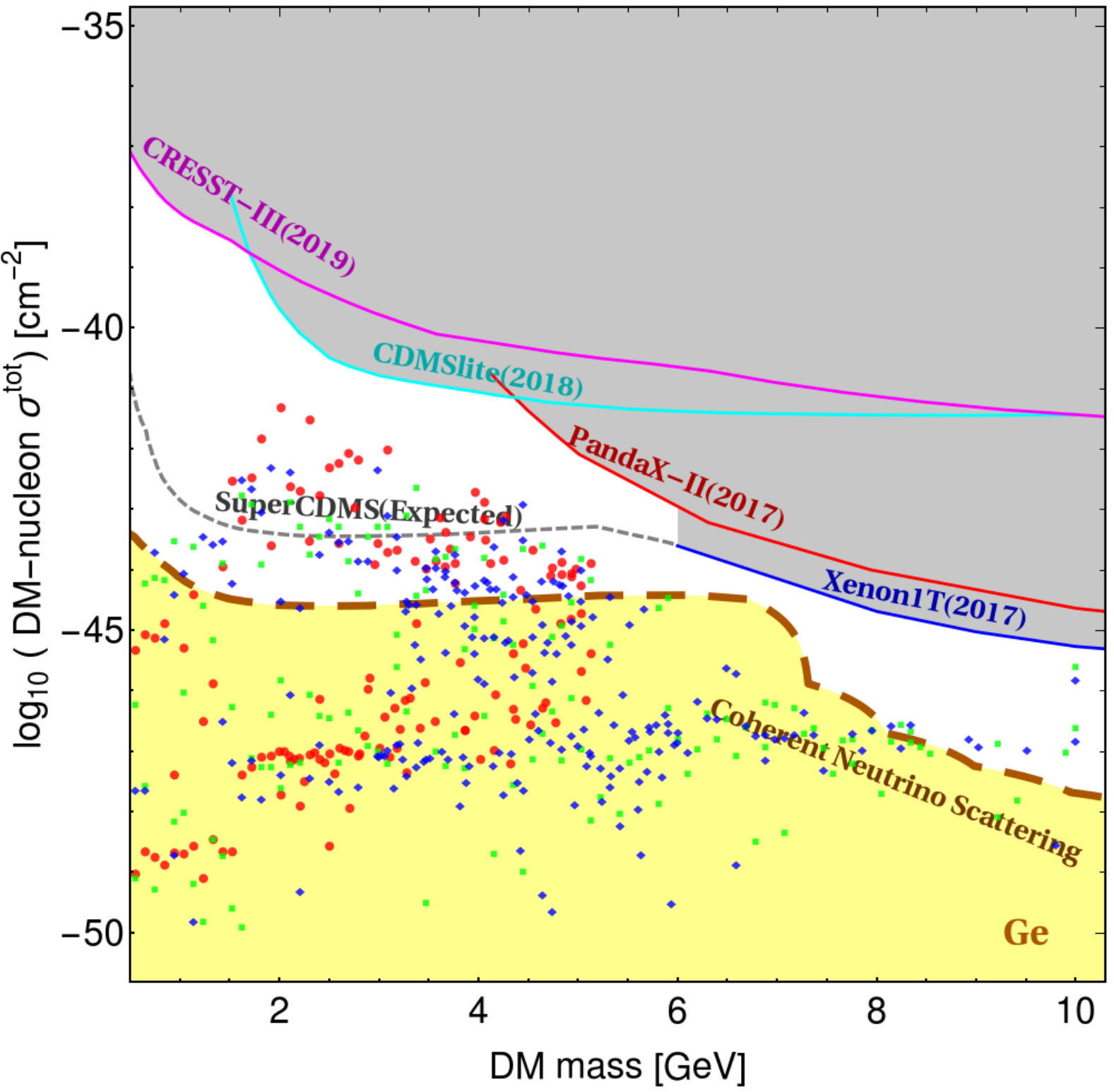}
$$
\caption{Experimental limits on the direct detection cross section $ \sigma $. The upper curves are obtained, from left to right, from the CRESST, CDMS, PandaX and Xenon1T experiments, and the expected sensitivity limit for the superCDMS experiment \cite{Agnese:2016cpb}; the coherent neutrino scattering regions are calculated for Xe (left) and Ge (right). For illustration we also include the cross sections corresponding to a selection of points on the boundary of the allowed region of parameter space, on the upper and lower boundaries of Fig. \ref{fig:MPhi-mix}(a) (green points), of Fig. \ref{fig:MPhi-mix}(c) (red points),  and of Fig. \ref{fig:MPhi-mix}(d) (blue points).}
\label{fig:DD_limits}
\end{figure}

In Fig.~\ref{fig:DD_limits} we plot the values of the direct-detection cross sections  for a selection of points on or close to the boundary of the allowed region of parameter space. The points are chosen only to illustrate that there is a region of parameter space within the sensitivity reach of SuperCDMS~\cite{Agnese:2016cpb}, but that this experiment cannot exclude the model; it is also worth noting that a (different)  region of parameter space will correspond to cross sections above the coherent neutrino scattering `floor'. Both these regions are significant in size: restricting the model to either (or both) would not require fine tuning.

\section{Conclusions}
\label{sec:conclusions}

 In this paper, we have considered an extension of the neutrino-portal DM scenario, introducing strong interactions to the dark sector via a $\ui_{\tt dark} $ local symmetry with its corresponding vector boson $V$. The dark sector consists of a scalar $ \Phi $, the dark photon $V$, and two almost degenerate fermions $ \Psi_\pm$, of opposite $\ui_{\tt dark} $ charges and which constitute the DM relics. We have also imposed a (softly broken) $\mathbb{Z}_{2}$ dark-charge symmetry that strongly suppresses $V$ mixings with the SM photon and $Z$, but still allows for the $V$ to decay into neutrinos with a sufficiently short lifetime, as required by phenomenology; this imposes mild constraints on the soft breaking parameter. These modifications to the model preserve the naturally small direct and indirect detection cross sections and the relatively large annihilation cross-sections without fine-tuning.
 
The core vs. cusp data in galaxies and clusters place a limit on the DM self-interaction cross sections, from which we derive limits on the strong interaction coupling and the mass of the $V$ boson. The presence of oppositely charged DM components opens the possibility that bound states are formed; if this occurs, and the formation of rate such bound states is sufficiently high the core vs. cusp problem would reappear as the interactions between the bound states will be weak (akin to the Van der Waals interactions). In this paper we took a conservative approach and simply required that the potential generated by the $V$ should not lead to $ \BS
$  bound states by assuming that these particles are sufficiently light, accordingly we have chosen $ \mfe \lesssim O( 10\, \gev) $ in our numerical calculations; we will return to the issue of bound state formation in a  future publication.  

The relic density constraint also imposes strong restrictions on the model. Specifically, a large dark photon coupling $g$ leads (cf. \cref{eq:sig0}) to under-production of DM, so the upper allowed values (see \cref{eq:mv.g.mfe,eq:mv.g.errors}) for this coupling are generally problematic. A more precise determination of the DM cross section as a function of velocity will provide a strict test of the viability of this model. In addition, $Z$ and $W$ data, impose important restrictions on the mixing angles $ \scal $ and Yukawa couplings $z$.

Other constraints on the model are milder. For example, the DM-nucleon cross section is naturally suppressed in this model (it is a one-loop effect), so that the direct-detection limits provide less significant in restrictions than in other models. We have not included constraints derived from neutrino oscillations because they are not precise enough to provide significant limits. The same applies to existing limits derived from the measurement of the muon anomalous magnetic moment, in this case, however, an improvement by one order of magnitude in the experimental sensitivity would provide useful constraints on this model. 

As with the original model \cite{Gonzalez-Macias:2016vxy}, the most distinct detection signature would come from the annihilation of $\Psi$'s into neutrinos, producing a monochromatic neutrino line from both the sun and the galactic halo; unfortunately,  current detection experiments have insufficient sensitivity to detect such a signal.

Also of  interest are the allowed values of the mass of the dark scalars, $ \mfe < \msc \lesssim 300\, \gev  $ (Fig. \ref{fig:MPhi-mix} $(a)$). The existence of this particle can be probed in principle by accurate measurements of the cosmological or astrophysical  neutrino flux, since it will exhibit a resonance at neutrino energy $ E_\nu\up{\tt res} = ( \msc^2 - \mfe^2)/(2 \mfe ) $ in the scattering of high-energy neutrinos off the ambient DM. Numerically, $  E_\nu\up{\tt res} \sim 3.7 \, \tev $ (roughly independent of $\mfe $) for the maximum allowed values of $ \msc  $ (upper boundary in the figure). Observation of this effect is challenging because the atmospheric neutrino flux is much larger at these energies.

The presence of a dark photon generates a significant change from the previous model \cite{Gonzalez-Macias:2016vxy}. The dark photons are long-lived and will decay into neutrinos; possible effects of these decays will be explored in a future publication.

\section*{Acknowledgments}
\label{sec:acknowledgments}

This work has been supported by the UC MEXUS-CONACYT collaborative grant CN-18-128. The authors would like to thank Hai-bo Yu for interesting and useful comments. J.M.L. acknowledges the University of California, Riverside, for its warm hospitality.

\bibliography{bibliography}{}
\bibliographystyle{JHEP}

\end{document}

%% file: mylatex.tex
\usepackage{bbm}  
\usepackage{amsmath}                                                

\newcommand{\nc}{\newcommand}
\nc{\beq}{\begin{equation}}  \nc{\eeq}{\end{equation}}
\nc{\bea}{\begin{eqnarray}}  \nc{\eea}{\end{eqnarray}}
\nc{\baa}{\begin{array}}     \nc{\eaa}{\end{array}}
\nc{\bit}{\begin{itemize}}   \nc{\eit}{\end{itemize}}
\nc{\ben}{\begin{enumerate}} \nc{\een}{\end{enumerate}}
\nc{\bce}{\begin{center}}    \nc{\ece}{\end{center}}
\nc{\bpm}{\begin{pmatrix}}   \nc{\epm}{\end{pmatrix}}
\nc{\bvt}{\begin{verbatim}}  \nc{\evt}{\end{verbatim}}
\nc{\bal}{\begin{align}}
\def\mcr{\nonumber\\[6pt]}

\def\half{\frac12}	

\def\to{\rightarrow}

\def\boldoverdot{\,{\raise6pt\hbox{\bf.}\!\!\!\!\>}}

\def\then{{\quad\Rightarrow\quad}}
\def\acal{{\cal A}}

\def\ccal{{\cal C}}

\def\fcal{{\cal F}}

\def\jcal{{\cal J}}

\def\lcal{{\cal L}}

\def\ncal{{\cal N}}

\def\scal{{\cal S}}

\def\qq{{\bf q}}

\def\ss{{\bf s}}

\def\kBB{{\mathbbm K}}

\def\mati{{\mathbbm1}}

\usepackage{bm}
\def\ssb{spontaneous symmetry breaking}
\def\vev{vacuum expectation value}

\def\tr{ \hbox{tr}}

\def\diag{\hbox{\diag}}
\def\sm{Standard Model}

\def\mev{\hbox{MeV}}
\def\gev{\hbox{GeV}}
\def\tev{\hbox{TeV}}

\def\cm{\hbox{cm}}

\def\gr{\hbox{gr}}

\def\vevof#1{\left\langle #1 \right\rangle}

\def\doubleundertext#1{
{\undertext{\vphantom{y}#1}}\par\nobreak\vskip-\the\baselineskip\vskip4pt%
\undertext{\hbox to 2in{}}}
\def\inbox#1{\vbox{\hrule\hbox{\vrule\kern5pt
     \vbox{\kern5pt#1\kern5pt}\kern5pt\vrule}\hrule}}
\def\sqr#1#2{{\vcenter{\hrule height.#2pt
      \hbox{\vrule width.#2pt height#1pt \kern#1pt
         \vrule width.#2pt}
      \hrule height.#2pt}}}

\def\today{\ifcase\month\or
  January\or February\or March\or April\or May\or June\or
  July\or August\or September\or October\or November\or December\fi
  \space\number\day, \number\year}
\def\pmb#1{\setbox0=\hbox{#1}%
  \kern-.025em\copy0\kern-\wd0
  \kern.05em\copy0\kern-\wd0
  \kern-.025em\raise.0433em\box0 }

\def\pmbb#1{\setbox0=\hbox{#1}%
  \kern-.02em\copy0\kern-\wd0
  \kern.04em\copy0\kern-\wd0
  \kern-.02em\raise.03464em\box0 }
\def\up#1{^{\left( #1 \right) }}
\def\inv#1{\frac1{#1}}
\def\su#1{{SU(#1)}}
\def\ui{U(1)}
\def\sumprime_#1{\setbox0=\hbox{$\scriptstyle{#1}$}
  \setbox2=\hbox{$\displaystyle{\sum}$}
  \setbox4=\hbox{${}'\mathsurround=0pt$}
  \dimen0=.5\wd0 \advance\dimen0 by-.5\wd2
  \ifdim\dimen0>0pt
  \ifdim\dimen0>\wd4 \kern\wd4 \else\kern\dimen0\fi\fi
\mathop{{\sum}'}_{\kern-\wd4 #1}}